\documentclass[aps,prb,twocolumn,showpacs,superscriptaddress,groupedaddress]{revtex4-1}
\usepackage{graphicx}
\usepackage{dcolumn}
\usepackage{bm}
\usepackage{amssymb}
\usepackage{amsmath}
\usepackage{subfigure}

\begin{document}

\title{Layer Antiferromagnetic State in Bilayer Graphene : A First-Principle Investigation}

\author{Yong Wang}
\affiliation{Department of Physics, The University of Hong Kong,
Hong Kong SAR, China}

\author{Hao Wang}
\affiliation{Department of Physics, The University of Hong Kong,
Hong Kong SAR, China}

\author{Jin-Hua Gao}
\email{jhgao1980@gmail.com} \affiliation{Department of Physics,
Huazhong University of Science and Technology, Wuhan, Hubei, China }
 \affiliation{Department of Physics,
The University of Hong Kong, Hong Kong SAR, China}

\author{Fu-Chun Zhang}
\affiliation{Department of Physics, The University of Hong Kong,
Hong Kong SAR, China}
\affiliation{Department of Physics, Zhejiang
University, Hangzhou, China}

\begin{abstract}
The ground state  of bilayer graphene is investigated by the
density functional calculations with local spin density
approximation. We find a ground state with layer antiferromagnetic
ordering, which has been suggested by former studies based on
simplified model. The calculations prove that the layer
antiferromagnetic state (LAF) is stable even if the remote hopping
and nonlocal Coulomb interaction are included. The gap of the LAF
state is about $1.8$~meV, comparable to the experimental value. The
surface magnetism in BLG is of the order of $10^{-2} \mu_B /nm^2 $.
\end{abstract}

\maketitle

Graphene, the star material nowadays, has shown many exotic physical
properties and promised great potentials in developing new
electronics devices\cite{rev1,rev2, rev3, rev4}. However, one major
roadblock for the practical applications of graphene in electronics
is the small on/off ratio due to its gapless ground state. One
alternative way to overcome the difficulty is exploiting the
AB-stacked bilayer graphene (BLG) instead, where a gap can be formed
and tuned by chemical doping or external electric
field\cite{doping,elecf}. For example, the unipolar transport has
been demonstrated in a field effect transistor based on
BLG\cite{unip}, and high frequency manipulation of the BLG quantum
dot has been realized\cite{hfreq}. Recently, the results of several
experiments on the ultraclean suspended BLG suggest that an
intrinsic gap may exist at the charge neutrality point, which is
attributed to the formation of certain ordered ground states due to
spontaneously symmetry broken\cite{exp1,exp2,exp3,exp4,exp5,exp6}. The
nature of the correlated ground state is still unclear and highly debated, and different
candidate states have been proposed theoretically, such as the layer
antiferromagnetic (LAF) state, quantum anomalous Hall state, quantum
spin Hall state, as well as a gapless nematic
state\cite{theo1,theo2,theo3,theo4,theo5,theo6,theo7,theo8}.

The LAF state is one of the most possible candidates for the
correlated ground state of BLG, which spontaneously breaks both the
spin rotational and sublattice symmetry. By quantum Monte Carlo,
renormalization group, and mean field methods based on a Hubbard
model, the LAF is shown to be stable over a wide range of parameter
range\cite{theo9,theo10,lafmf}. However, to determine the correlated
ground state of BLG, two important factors have been ignored in a
simple Hubbard model, which only focuses on the local Coulomb
interaction. One is the influence of other components of the Coulomb
interaction beyond the Hubbard U term, i.e. the long or short range
nonlocal Coulomb interaction. Although these terms are much smaller
than the on-site Coulomb interaction, some recent theoretical works
indicate that these terms do affect the correlated ground state of
BLG\cite{theo11,theo12}. The other is effect of the remote hopping
terms, which essentially modifies the parabolic feature of the
energy bands near the Fermi level at high symmetry points.
Furthermore, the choice of the model parameters also strongly
influences the calculating results.

In this work, we study the ground state of BLG  via density
functional theory (DFT) calculations with local spin density
approximation (LSDA). The results clearly indicate an insulating LAF
ground state with an energy gap about $1.8$ meV, which is of the
same order of magnitude of the experimental value, i.e about $2\sim 3$
meV\cite{exp4,exp5,exp6}. This is different from the former DFT
calculations on BLG in which a gapless ground state is
predicted if the spin degree freedom is not considered\cite{DFTgraph}.
We show that DFT calculation is helpful to investigate the correlated ground state of BLG,
especially the LAF state. Most importantly, we prove that the LAF  state
is stable in BLG even in the presence of nonlocal Coulomb
interaction and remote hopping, which are naturally included in DFT
calculation. It provides an essential support to the LAF ground
state. Furthermore, compared with other theoretical methods, the
first principles calculations give a more quantitative description
about the LAF ground state of BLG. The physical quantities which can
be detected in experiments, e.g. spin distribution between sites and
layers, are given in a more accurate way without empirical
parameters.

\begin{figure}[htp]
\includegraphics[scale=0.15,clip]{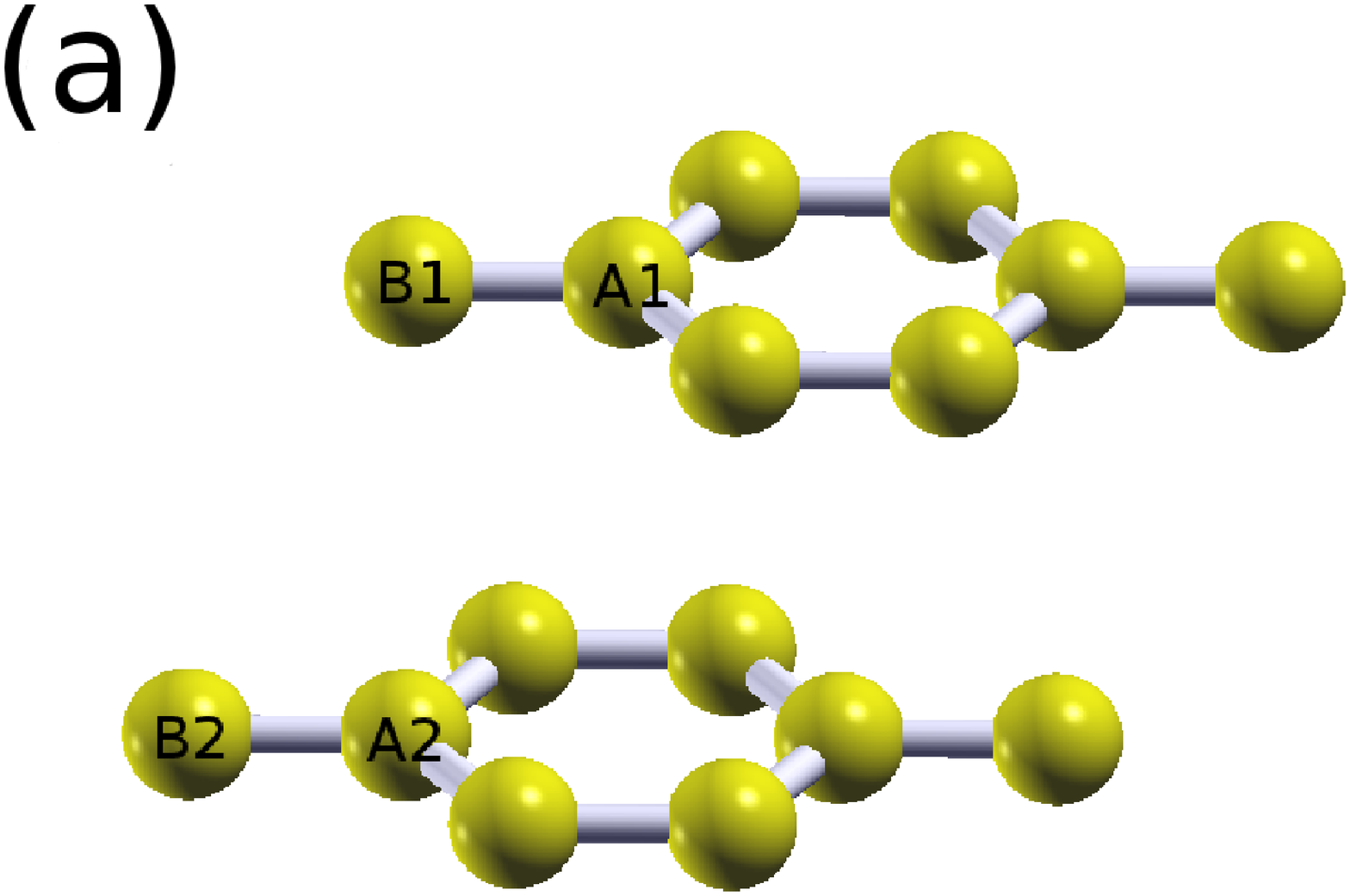}
\includegraphics[scale=0.15,clip]{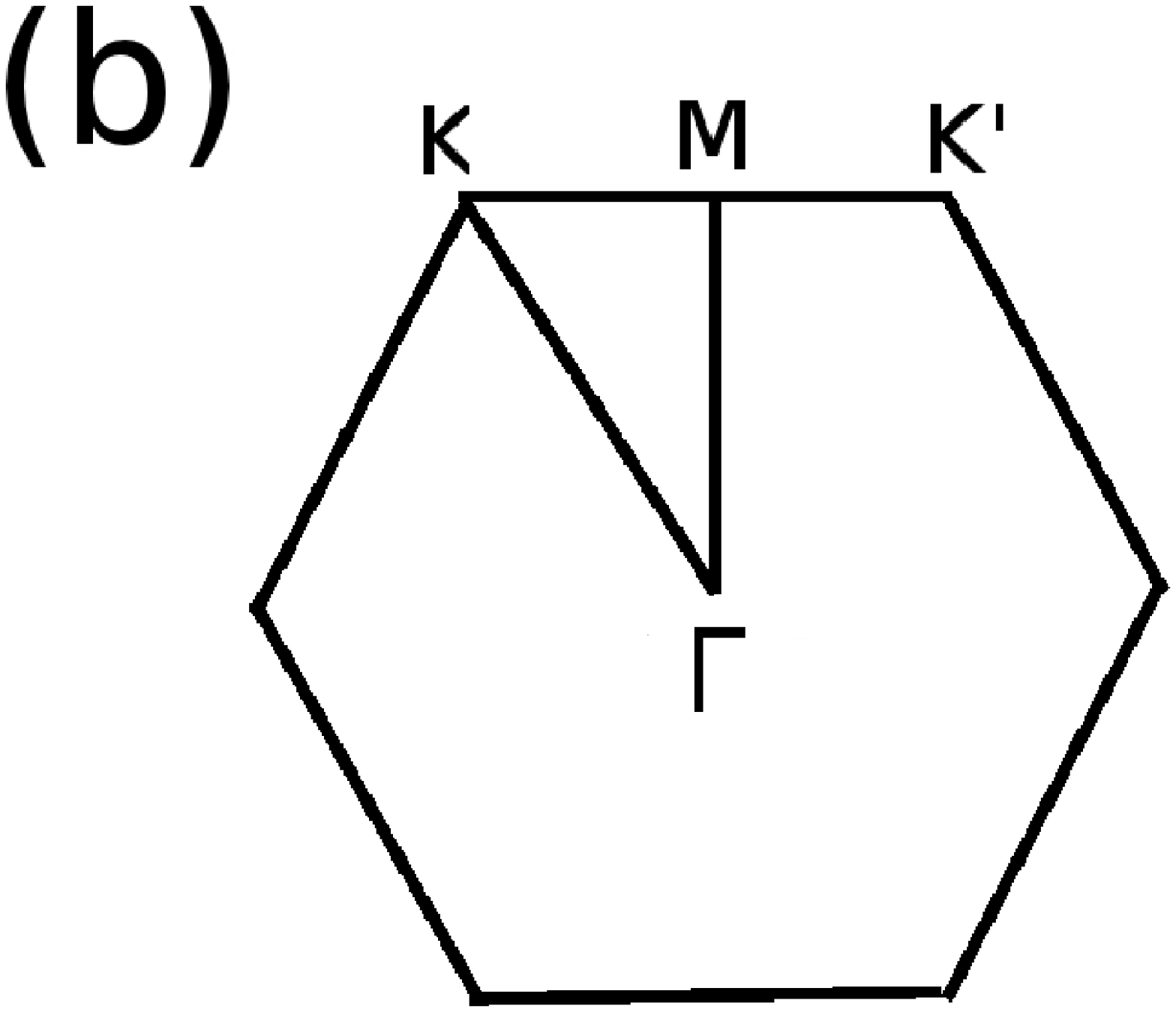}
\caption{(Color online) (a) The geometry structure of a $2\times2$
supercell of BLG. The four carbon atoms in the primitive cell are
denoted as A1, B1, A2, and B2 respectively; (b) Brillouin zone (BZ)
of BLG. $\Gamma$, $M$, $K$, and $K'$ denote the high symmetry points
in BZ. }\label{geom}
\end{figure}

The DFT calculations of the electron structures for BLG are
performed with the ABINIT software package\cite{ABinit}. There are
four carbon atoms in the primitive cell of the AB-stacking BLG, and
the geometry structure of a $2 \times 2$ supercell is shown in
Fig.~\ref{geom}~(a). In each layer, there are two sublattice, i.e.
sublattice  $A$ and $B$. We use the experimental value of the layer separation 3.35~\AA~,
since the van der Waals interactions can not be captured by DFT calculation\cite{ABCmac}.
A 23.4~\AA~vaccum layer is used to separate the BLGs in the
calculations with periodic boundary condition. The cutoff energy for
the plane wave basis set is chosen as 40~Ha, and the
Troullier-Martins (TM) norm-conserving pseudopotential for the
carbon element was exploited. The k points for sampling the
Brillouin zone (BZ) is generated by a $60\times60$ Monkhorst-Pack
(MP) grid. In order to investigate the possible spin order, the LSDA
is exploited to the exchange-correlation functional, and the
converge criteria for the total energy difference is $10^{-10}$~Ha.

\begin{figure}[htp]
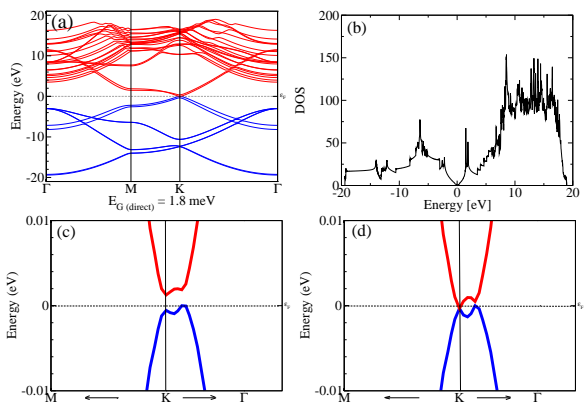

\includegraphics[scale=0.15,clip]{band.eps}
\includegraphics[scale=0.15,clip]{DOS.eps}
\includegraphics[scale=0.15,clip]{band2.eps}
\includegraphics[scale=0.15,clip]{band0.eps}
\caption{(Color online) (a) Band structure of BLG from DFT+LSDA
calculations; (b) DOS of BLG from DFT+LSDA calculations; (c) Fine
band structure of BLG around Dirac point $K$ from DFT+LSDA
calculations; (d) Fine band structure of BLG around Dirac point $K$
from DFT+LDA calculations.}\label{banddos}
\end{figure}

The band structures and density of states (DOS) of BLG from the
DFT+LSDA calculations are shown in Fig.~\ref{banddos}~(a) and (b)
respectively. A fine band structure around the Dirac point $K$ is
shown in Fig.~\ref{banddos}~(c), where we present our most
encouraging result, a minor band gap opened around the Dirac point
$K$. The calculated band gap $E_{gap} \approx 1.8$ meV is
comparable to the experimental values $2\sim 3$~meV\cite{exp4,exp5,exp6},
considering that LSDA calculations usually underestimate the band
gap. For comparison, we also perform the DFT+LDA calculation without
including the spin degrees of freedom.  In Fig.~\ref{banddos}~(d),
we show the fine band structure from this calculation, which
reproduces previous studies\cite{DFTgraph}. No energy gap is found
in this case. We see the trigonal warping in the energy band, which results from the remote hopping.
Note that, for trigonal warping, there are  three additional touching points at the Fermi level near K point in addition to the Dirac point. But in Fig. ~\ref{banddos}~(d) we only show one of the three touching points, since the other two are not on the high symmetry line. As shown in Fig. ~\ref{banddos}~(c), these degeneracies
are all lifted if we include the spin degree of freedom in the
calculation.

\begin{figure}[htp]
\includegraphics[scale=0.16,clip]{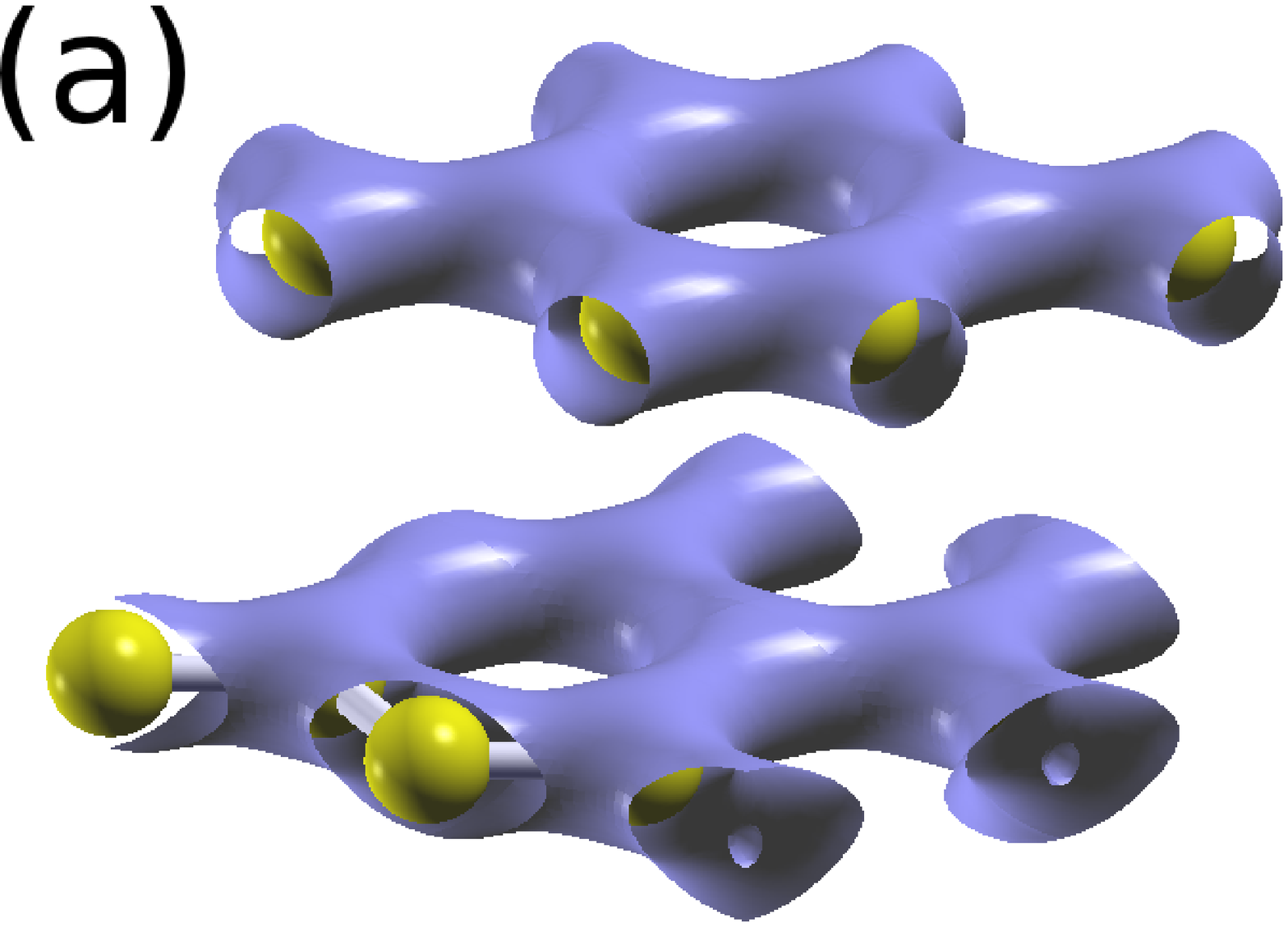}
\includegraphics[scale=0.16,clip]{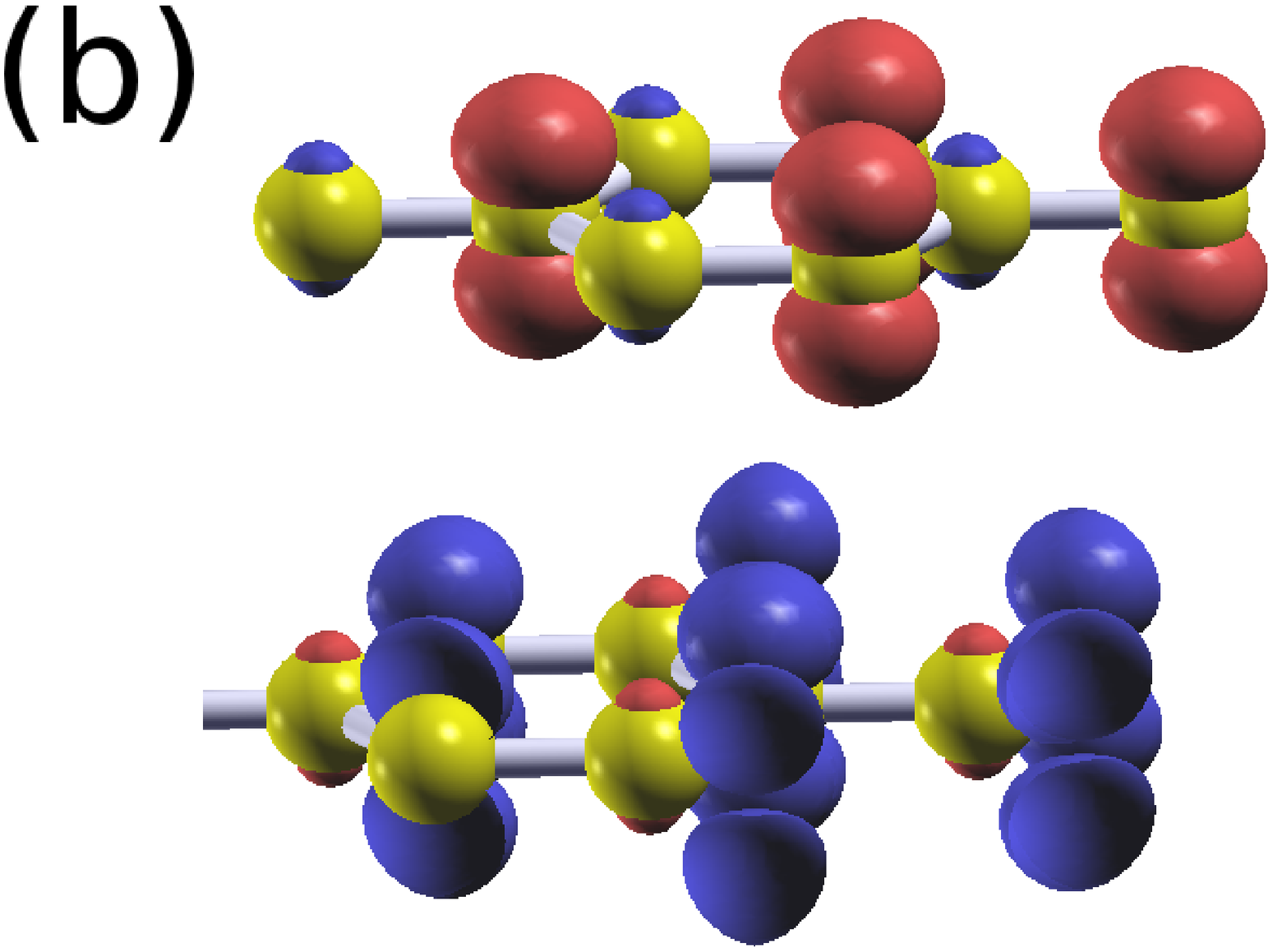}
\caption{(Color online) (a) Isosurface of the charge distribution
with isovalue 0.15; (b) Isosurface of the spin polarization
distribution with isovalue $2.5\times10^{-5}$(red) and
-$2.5\times10^{-5}$(blue).}\label{charspin}
\end{figure}

In order to identify the nature of the ground state of BLG further,
we analyze its charge density and spin-polarization distributions.
The isosurface of the charge density is illustrated in Fig.~\ref{charspin}~(a), which is
basically extended along the C-C bonds in each layer. We estimate
the charge on each atom using the Hirshfeld method\cite{HirF}.
Calculations with LDA and LSDA give the same charge distributions.
It implies that the energy gap in LSDA calculation has nothing to do
with the charge redistribution. We then check the spin distribution.
The calculation with LDA does not give any spin ordering, since the
spin degree of freedom has been ignored in this case. But in the
calculation with LSDA, we find a special spin ordering, i.e. layer
antiferromagnetic order. The spin ordering is shown clearly in Fig.
~\ref{charspin}~(b). The spin polarization is mainly localized around
the A1 and B2 atoms. By the Hirshfeld method, we get the spin
polarization around A1 atom is about $5.5 \times 10^{-4}$, and that around B2 is
$-5.5 \times 10^{-4}$. Note that, for spin polarization, we mean
$n_{\uparrow}-n_{\downarrow}$ where $n_{\uparrow}$ ($n_{\downarrow}$)
is the charge number with up spin (down spin). For the A2 and B1
atoms, the spin polarization is very tiny, and at least one order of magnitude smaller.
We see that the spin ordering between two nearest neighbor sites are all
antiferromagnetic. The numerical results show that the net spin in
each layer is nonzero but that of the whole system is zero. The spin
structure in each layer is antiferrimagnetic and the spin
polarization of two layers are of opposite sign. That is just the
layer antiferromagnetic state predicted by former mean field
studies. In Ref. 33, considering the experimental value of the gap 2
meV, self-consistent mean field calculation gives similar spin
ordering, and the largest spin polarization on one atom is of the
order of $10^{-4}$, which is in qualitatively or semi-quantitatively
agreement with our DFT results. Because that the remote hopping and
nonlocal Coulomb interaction are naturally included in DFT
calculation, the LAF state we found here indicate that the LAF state
is stable even in the presence of remote hopping and nonlocal
Coulomb interaction. It offers an essential support to former mean
field studies.   We emphasize here that DFT calculation is the best way so far to investigate the influences of remote hopping and nonlocal Coulomb interaction. Meanwhile, our DFT calculation shows that the
surface magnetism is about $10^{-2} \mu_B/nm^2$, which can be detected in experiment by spin-polarized
scanning tunneling microscopy. We note that the surface magnetism has been reported in trilayer and 8-layer graphene
systems\cite{8layer,8layer2,trilayer}.

\begin{figure}[htp]
\includegraphics[scale=0.2,clip]{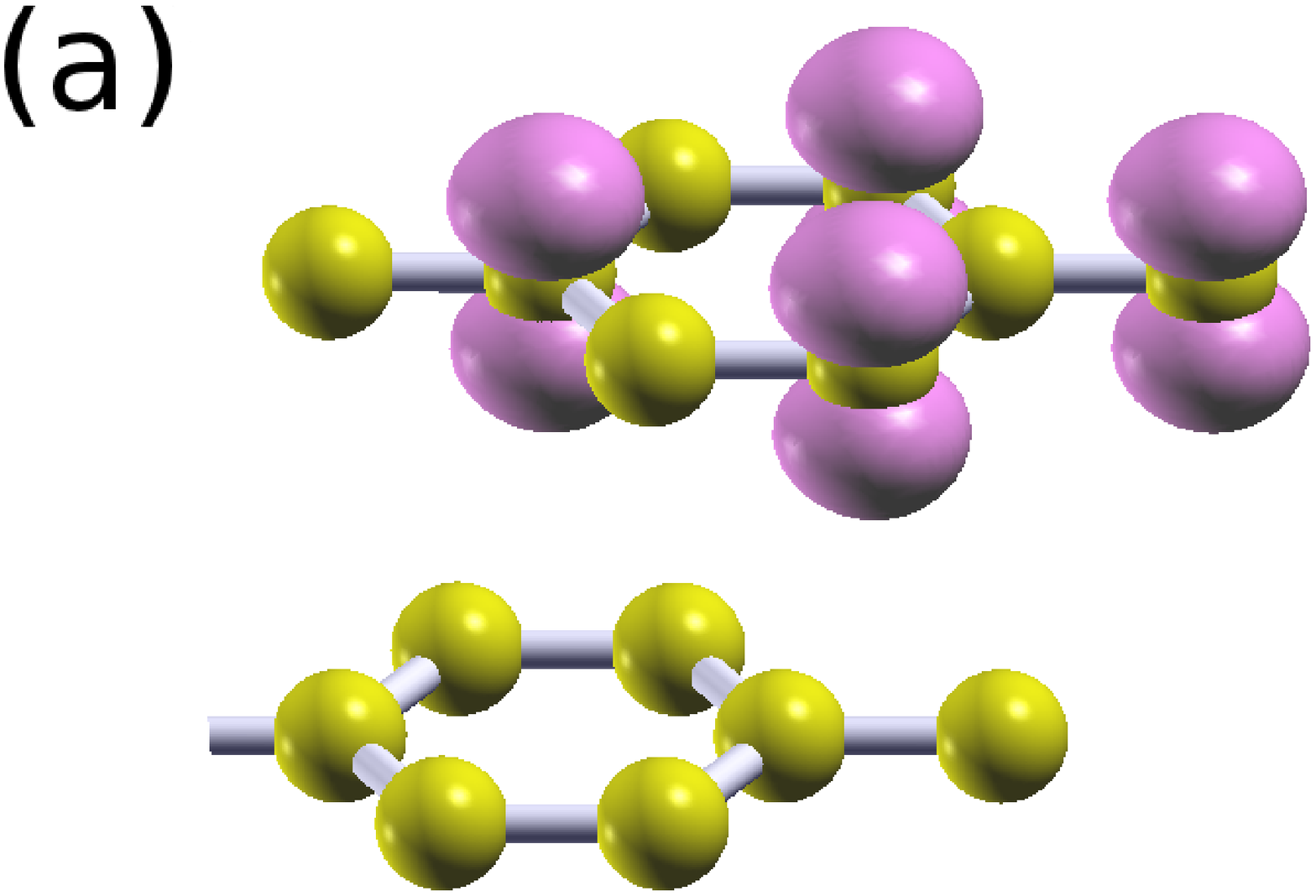}
\includegraphics[scale=0.2,clip]{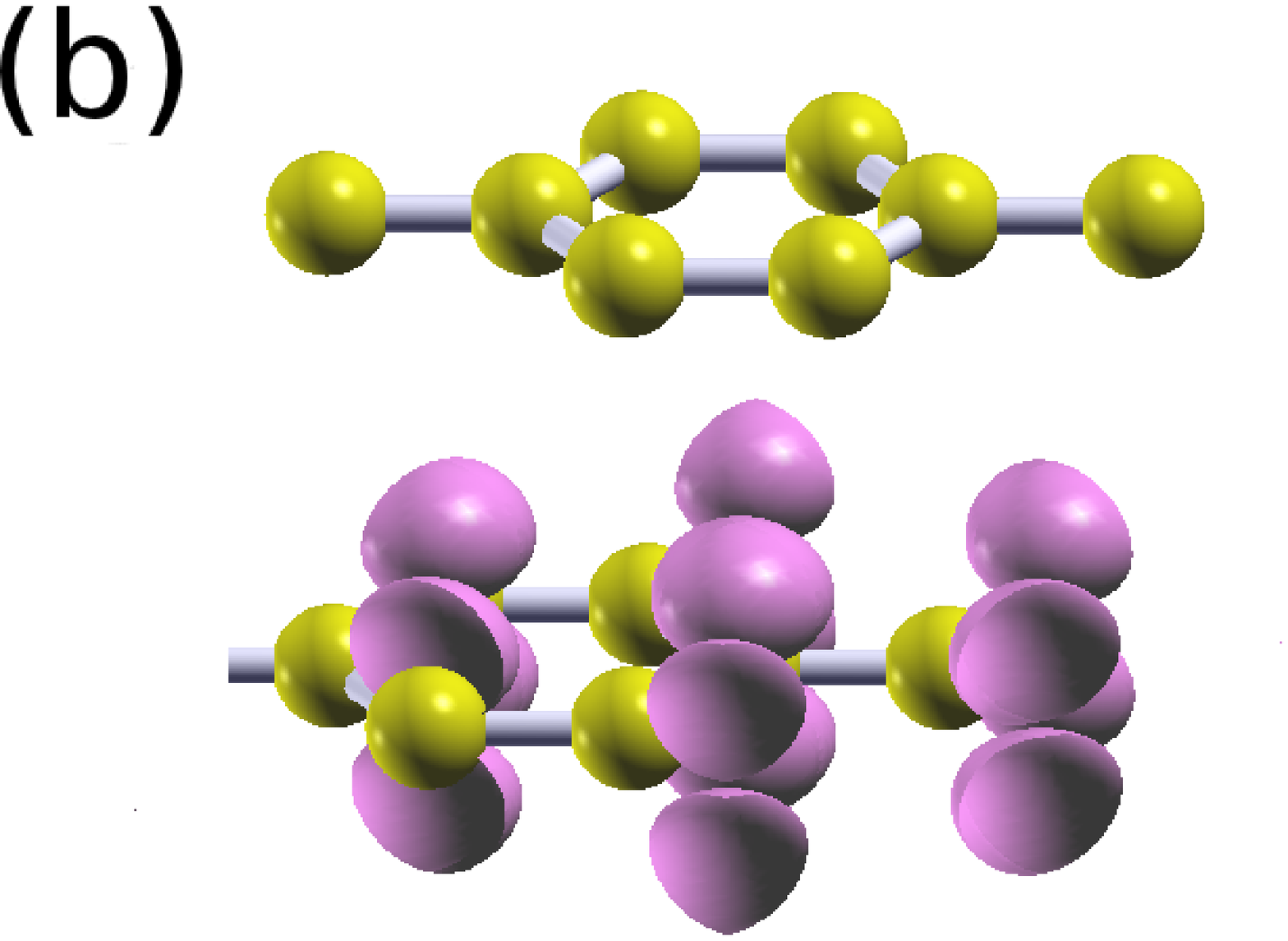}
\caption{(Color online) Isosurfaces of the norm of the wavefunction
(spin up) for  states at $K$ point. (a) The  state on valence band;
(b) the state on conductance band. The isovalue is 30
here.}\label{wave}
\end{figure}

Another confirmation of the LAF state is its peculiar wave function
near the Fermi level. In noninteracting case, the low energy states of BLG has a
pseudospin symmetry, i.e. the layer symmetry, in addition to the normal spin
and valley symmetry\cite{theo2}. As mentioned before,
electron-electron interaction can spontaneously break some symmetry
and induce several possible correlated ground states. In LAF state,
layer symmetry is broken for each spin direction, because the two layers have opposite spin
polarization. In other words, for up spin, the states on conductance
band near the Fermi level are localized on one layer, while that
on the valence band are on the other. For down spin, the layer
dependence is inverted. These features of the wave function of LAF
state have been demonstrated clearly in former studies based on
simplified model\cite{theo11,bilayerMF}. Our first principles results of
the BLG wave function also have these features. Taking one spin
direction for example (saying up spin), we plot the wave functions
for both conductance and valence bands at $K$ point in Fig.~\ref{wave}~(a) and (b).
As we expected, the state on valence (conductance) band is on the top (bottom) layer. However, the wave
function for the states a little away from the Fermi level does not
have such spin and layer dependent distribution. Actually, the wave
functions for these states spread uniformly among the two layers.
The results here confirm that the electron-electron
interaction in BLG system only influence the low energy states very
close to the Fermi level. And only these states have the spin
ordering.

In summary, the first principles calculations have been performed to
investigate the ground state of BLG. The LAF ground state is found
in the LDSA calculations and a reasonable energy gap compared with
the experimental results is obtained. This calculation proves that
the LAF state in BLG is stable in the presence of remote hopping and
nonlocal Coulomb interaction. Our calculations also give the values
of some key physical quantities of the LAF states. The largest spin
polarization on one atom is about $5.5 \times 10^{-4}$. The surface
magnetism is of the order of $10^{-2} \mu_B/nm^2$.

We acknowledge some of financial support from HK- SAR RGC Grant No.
HKU 701010, CRF Grant No. HKU 707010 and the Hong Kong University
Grant Council (AoE/P-04/08). J.H.G is supported by the National
Natural Science Foundation of China (Project No.11274129).


\begin{thebibliography}{99}
\bibitem{rev1} A. H. Castro Neto, F. Guinea, N. M. R. Peres
, K. S. Novoselov, and A. K. Geim, Rev. Mod. Phys. \textbf{81}, 109
(2009).
\bibitem{rev2} S. Das Sarma, S. Adam, E. H. Hwang, and E. Rossi, Rev. Mod. Phys. \textbf{83}, 407 (2011).
\bibitem{rev3} M. O. Goerbig, Rev. Mod. Phys. \textbf{83}, 1193 (2011).
\bibitem{rev4} V. N. Kotov, B. Uchoa, V. M. Pereira, F. Guinea, A. H. C. Neto, Rev. Mod. Phys. \textbf{84}, 1067 (2012).
\bibitem{doping} T. Ohta, Aaron Bostwick, T. Seyller, K. Horn, and E. Rotenberg, Science, \textbf{313}, 951 (2006).
\bibitem{elecf} Y. Zhang, T.-T. Tang, C. Girit, Z. Hao, M. C. Martin, A. Zettl, M. F. Crommie, Y. R. Shen, and F. Wang, Nature, \textbf{459}, 820 (2009).
\bibitem{unip} H. Miyazaki, S.-L. Li, S. Nakaharai, and K. Tsukagoshi, Appl. Phys. Lett. \textbf{100}, 163115 (2012).
\bibitem{hfreq} S. Droscher, J. Guttinger, T. Mathis, B. Batlogg, T. Ihn, and K. Ensslin, Appl. Phys. Lett. \textbf{101}, 043107 (2012).
\bibitem{exp1} B. E. Feldman, J. Martin, and A. Yacoby, Nature Phys. \textbf{5}, 889 (2009).
\bibitem{exp2} R. T. Weitz, M. T. Allen, B. E. Feldman, J. Martin, and A. Yacoby, Science, \textbf{330}, 812 (2010).
\bibitem{exp3} A. S. Mayorov, D. C. Elias, M. Mucha-Kruczynski, R. V. Gorbachev, T. Tudorovskiy, A. Zhukov, S. V. Morozov, M. I. Katsnelson, V. I. Fal’ko, A. K. Geim, and K. S. Novoselov, Science, \textbf{333}, 860 (2011).
\bibitem{exp4} J. Velasco Jr, L. Jing, W. Bao, Y. Lee, P. Kratz,
V. Aji, M. Bockrath, C. N. Lau, C. Varma, R. Stillwell, D. Smirnov,
F. Zhang, J. Jung, and A. H. MacDonald, Nature Nanotech. \textbf{7},
156 (2012).
\bibitem{exp5} W. Bao, J. Velasco, Jr., F. Zhang, L. Jing, B. Standley, D. Smirnov, M. Bockrath, A. H. MacDonald, and C. N. Lau, Proc. Natl. Acad. Sci. U.S.A. \textbf{109}, 10802 (2012).
\bibitem{exp6} F. Freitag, J. Trbovic, M. Weiss, and C. Sch\"{o}nenberger, Phys. Rev. Lett. \textbf{108}, 076602 (2012).
\bibitem{theo1} J. Nilsson, A. H. Castro Neto, N. M. R. Peres, and F. Guinea, Phys. Rev. B \textbf{73}, 214418 (2006).
\bibitem{theo2} H. Min, G. Borghi, M. Polini, and A. H. MacDonald, Phys. Rev. B \textbf{77}, 041407 (2008).
\bibitem{theo3} O. Vafek and K. Yang, Phys. Rev. B \textbf{81}, 041401(R)(2010).
\bibitem{theo4} Y. Lemonik, I. L. Aleiner, C. Toke, and V. I. Fal’ko, Phys. Rev. B \textbf{82}, 201408 (2010).
\bibitem{theo5} R. Nandkishore and L. Levitov, Phys. Rev. B \textbf{82}, 115124 (2010).
\bibitem{theo6} F. Zhang, H. Min, M. Polini, and A. H. MacDonald, Phys. Rev. B \textbf{81}, 041402(R) (2010).
\bibitem{theo7} F. Zhang, J. Jung, G. A. Fiete, Q. Niu, and A. H. MacDonald, Phys. Rev. Lett. \textbf{106}, 156801 (2011).
\bibitem{theo8} Y. Lemonik, I. L. Aleiner, and V. I. Fal’ko, Phys. Rev. B \textbf{85}, 245451 (2012).
\bibitem{theo9} T. C. Lang, Z. Y. Meng, M. M. Scherer, S. Uebelacker, F. F. Assaad, A. Muramatsu, C. Honerkamp, and S. Wessel, Phys. Rev. Lett. \textbf{109}, 126402 (2012).
\bibitem{theo10} M. M. Scherer, S. Uebelacker, and C. Honerkamp, Phys. Rev. B \textbf{85}, 235408 (2012).
\bibitem{lafmf} M. Kharitonov, Phys. Rev. B \textbf{86}, 195435 (2012).
\bibitem{theo11} F.Zhang, A. H. MacDonald, Phys. Rev. Lett. \textbf{108}, 186804 (2012); F. Zhang, H. Min, A. H. MacDonald, Phys. Rev. B \textbf{86}, 155128 (2012).
\bibitem{theo12} V. Cvetkovic, R. E. Throckmorton, and O. Vafek, Phys. Rev. B 86, 075467 (2012).
\bibitem{DFTgraph} S. Latil and L. Henrard, Phys. Rev. Lett. \textbf{97}, 036803 (2006).
\bibitem{ABinit}  X. Gonze, B. Amadon, P.-M. Anglade, J.-M. Beuken, F. Bottin, P. Boulanger, F. Bruneval, D. Caliste, R. Caracas, M. Cote, T. Deutsch, L. Genovese, Ph. Ghosez, M. Giantomassi, S. Goedecker, D.R. Hamann, P. Hermet, F. Jollet, G. Jomard, S. Leroux, M. Mancini, S. Mazevet,
 M.J.T. Oliveira, G. Onida, Y. Pouillon, T. Rangel, G.-M. Rignanese, D. Sangalli, R. Shaltaf, M. Torrent, M.J. Verstraete, G. Zerah, J. W. Zwanziger, Computer Phys. Comm. \textbf{180}, 2582 (2009).
\bibitem{ABCmac} F. Zhang, B. Sahu, H. Min, and A. H. MacDonald, Phys. Rev. B \textbf{82}, 035409 (2010).

\bibitem{twarp} E. McCann and V. I. Fal'ko, Phys. Rev. Lett. \textbf{96}, 086805 (2006).
\bibitem{HirF} F. L. Hirshfeld, Theor. Chem. Acta  \textbf{44}, 129 (1977).
\bibitem{bilayerMF} Jie Yuan, Jin-Hua Gao, Dong-hui Xu, Hao Wang, Yi Zhou and Fu-Chun Zhang, unpublished.
\bibitem{8layer} M. Otani, M. Koshino, Y. Takagi, and S. Okada, Phys. Rev. B \textbf{81}, 161403(R) (2010).
\bibitem{8layer2} M. Otani, Y. Takagi, M. Koshino, and S. Okada, Appl. Phys. Lett. \textbf{96}, 242504 (2010).
\bibitem{trilayer} D.-H. Xu, J. Yuan, Z.-J. Yao, Y. Zhou, J.-H. Gao, F.-C. Zhang,
Phys. Rev. B \textbf{86}, 201404 (R) (2012).
\bibitem{wang:2012} Hao Wang, J.-H. Gao, and F.-C. Zhang, arXiv:1210.7590.
\end{thebibliography}
\end{document}